\documentclass[12pt]{article}
\usepackage[pdftex]{graphicx} 
\usepackage{epsfig}
\usepackage{comment}
\usepackage{latexsym}
\usepackage{hyperref}
\usepackage{amsmath}
\usepackage{color}

\newcommand{\mysquare}[0]{\raise-.2ex\hbox{{\Large$\Box$}}}
\def\lsim{\mathrel{\rlap {\raise.5ex\hbox{$ < $}}
{\lower.5ex\hbox{$\sim$}}}}
\def\gsim{\mathrel{\rlap {\raise.5ex\hbox{$ > $}}
{\lower.5ex\hbox{$\sim$}}}} \topmargin -1.5cm \textheight=22.5cm \textwidth=16.5cm
\catcode`\@=11
\newcount\hour
\newcount\minute
\newtoks\amorpm
\hour=\time\divide\hour by60 \minute=\time{\multiply\hour by60
\global\advance\minute by-\hour}
\edef\standardtime{{\ifnum\hour<12 \global\amorpm={am}%
        \else\global\amorpm={pm}\advance\hour by-12 \fi
        \ifnum\hour=0 \hour=12 \fi
        \number\hour:\ifnum\minute<10 0\fi\number\minute\the\amorpm}}
\edef\militarytime{\number\hour:\ifnum\minute<10 0\fi\number\minute}
\def\draftlabel#1{{\@bsphack\if@filesw {\let\thepage\relax
   \xdef\@gtempa{\write\@auxout{\string
      \newlabel{#1}{{\@currentlabel}{\thepage}}}}}\@gtempa
   \if@nobreak \ifvmode\nobreak\fi\fi\fi\@esphack}
        \gdef\@eqnlabel{#1}}
\def\@eqnlabel{}
\def\@vacuum{}
\def\draftmarginnote#1{\marginpar{\raggedright\scriptsize\tt#1}}
\def\draft{\oddsidemargin -.2truein
        \def\@oddfoot{\sl preliminary draft \hfil
        \rm\thepage\hfil\sl\today\quad\militarytime}
        \let\@evenfoot\@oddfoot \overfullrule 3pt
        \let\label=\draftlabel
        \let\marginnote=\draftmarginnote
   \def\@eqnnum{(\theequation)\rlap{\k

 ern\marginparsep\tt\@eqnlabel}%
\global\let\@eqnlabel\@vacuum}  }

\newcommand{\be}[0]{\begin{equation}}
\newcommand{\ee}[0]{\end{equation}}
\newcommand{\ba}[0]{\begin{eqnarray}}
\newcommand{\ea}[0]{\end{eqnarray}}
\newcommand{\dis}{\displaystyle}

\def\bs{\begin{subequations}}
\def\es{\end{subequations}}

\def\thebibliography#1{%
\vskip 0.5cm \centerline{\bf \Large References}
\list{%
[\arabic{enumi}]}{\settowidth\labelwidth{[#1]} \leftmargin\labelwidth
\advance\leftmargin\labelsep
\usecounter{enumi}}
\def\newblock{\hskip .11em plus .33em minus .07em}
\sloppy\clubpenalty4000\widowpenalty4000 \sfcode`\.=1000\relax}

\renewcommand{\theequation}{\arabic{section}.\arabic{equation}}

\renewcommand{\section}{\setcounter{equation}{0}\@startsection
{section}{1}{0mm}{-\baselineskip}{0.5\baselineskip} {\normalfont\Large\bfseries}}

\renewcommand{\subsection}{\@startsection
{subsection}{2}{0mm}{-\baselineskip}{0.5\baselineskip} {\normalfont\large\bfseries}}

\renewcommand{\subsubsection}{\@startsection
{subsubsection}{3}{0mm}{-\baselineskip}{0.5\baselineskip}
{\normalfont\normalsize\slshape}}

\usepackage{amssymb}
\usepackage{amsthm}
\usepackage{amsmath}
\usepackage{amssymb,amsfonts}
\usepackage{graphicx}
\usepackage{cite}

\newcommand{\ie}{{\em i.e. }}

\newcommand{\R}{\mathbb{R}}

\newcommand{\sign}{\mbox{sign}}

\renewcommand{\O}{{\cal O}}

\newcommand{\abs}{|}
\newcommand{\where}{\mbox{where}}

\renewcommand{\and}{\mbox{and}}
\newcommand{\esp}{\!\!\!\phantom{\Bigg\abs}}
\newcommand{\desp}{\!\!\! \phantom{\underset{\hat \abs}{\Big\abs}}}

\renewcommand{\v}{\vec}

\newcommand{\N}{{\cal N}}
\renewcommand{\S}{{\cal S}}

\newcommand{\B}{{\cal B}}

\newcommand{\C}{{\cal C}}

\begin{document}
\begin{titlepage}
\begin{flushright}
LPTENS--11/42,
CPHT--RR095.1011,
November  2011
\vspace{-.0cm}
\end{flushright}
\begin{centering}
{\bf \Large S-brane to thermal non-singular string cosmology}

\vspace{5mm}

 {\bf Costas Kounnas$^{1}$, Herv\'e Partouche$^2$ and Nicolaos Toumbas$^3$}

 \vspace{1mm}

$^1$ Laboratoire de Physique Th\'eorique,
Ecole Normale Sup\'erieure,$^\dag$ \\
24 rue Lhomond, F--75231 Paris cedex 05, France\\
{\em  Costas.Kounnas@lpt.ens.fr}

$^2$  {Centre de Physique Th\'eorique, Ecole Polytechnique,$^\ddag$\\
F--91128 Palaiseau cedex, France\\
{\em herve.partouche@cpht.polytechnique.fr}}

$^3$  Department of Physics, University of Cyprus,\\
Nicosia 1678, Cyprus.\\
{\em nick@ucy.ac.cy}

\end{centering}
\vspace{0.1cm}
$~$\\
\centerline{\bf\Large Abstract}\\
\vspace{-0.2cm}

\begin{quote}
We present a new class of non-singular string cosmologies in $d$ space-time dimensions. At very early times, $\tau\ll \tau_c$, the Universe is described by a flat $\sigma-$model metric, a constant maximal temperature $T_c$ and super-weak string interactions, $g_{\rm str}\ll 1$. During the evolution, the metric remains flat up to $\tau_c$, while the string coupling grows and reaches a critical value $g^*_{\rm str}$ at $\tau_c$. This phase is characterized by a uniform temporal distribution of spacelike branes. At later times, $\tau >\tau_c$, the Universe enters in a new phase of expansion, with radiation. The string coupling decreases due to the dilaton motion and asymptotes to a constant for $\tau\gg \tau_c$. Throughout the evolution, the string coupling remains smaller than $g^*_{\rm str}$. In the Einstein frame, the cosmologies describe bouncing Universes, where two distinct phases are connected at $\tau_c$. In the initial contracting phase, the evolution of the scale factor is identical to that of a negatively curved Universe filled with radiation. 
At later times, the Universe enters in an expanding thermal phase with a running dilaton. Explicit examples are presented in a large class of thermal $(4,0)$ type II superstring vacua, with non-trivial ``gravito-magnetic'' fluxes.

\end{quote}

\vspace{3pt} \vfill \hrule width 6.7cm \vskip.1mm{\small \small \small
  \noindent
   $^\dag$\ Unit{\'e} mixte  du CNRS et de l'Ecole Normale Sup{\'e}rieure associ\'ee \`a
l'Universit\'e Pierre et Marie Curie (Paris 6), UMR 8549.\\
$^\ddag$\ Unit{\'e} mixte du CNRS et de l'Ecole Polytechnique,
UMR 7644.}\\

\end{titlepage}
\newpage
\setcounter{footnote}{0}
\renewcommand{\thefootnote}{\arabic{footnote}}
 \setlength{\baselineskip}{.7cm} \setlength{\parskip}{.2cm}

\setcounter{section}{0}


\section{Introduction}
In string theory, one expects a drastically different cosmological picture to emerge, as compared to the conventional field theory scenario \cite{HE}, especially at  
very early cosmological times, where new stringy degrees of freedom become relevant, dominating the high temperature and high curvature regimes \cite{CosmoPheno,GV,GGV}. In these extreme situations, purely stringy phenomena occur, which do not admit conventional field theory descriptions \cite{CosmoTopologyChange}. String oscillators and winding states become relevant around the Hagedorn temperature $T_H$, before the onset of curvature singularities, 
and drive a phase transition towards a non-trivial thermal vacuum 
\cite{CosmoPheno,GV,AW,RK,AKADK,BR,DLS,Chaud,DL,HotstringsCosmo}. 

Non-singular cosmological solutions in various space-time dimensions were recently found, based on a mechanism which resolves the Hagedorn instabilities of strings at finite temperature \cite{FKPT,dNonSingular}. The key ingredients of this mechanism were shown to be generic in a large class of initially $\N_4=(4,0)$ superstring models, where finite temperature is introduced along with special {\it ``gravito-magnetic fluxes"}. The fluxes modify the thermal vacuum by injecting into it non-trivial momentum and winding charges, lifting the Hagedorn instabilities of the canonical ensemble \cite{akpt,massivesusy,FKT,FKPT,dNonSingular}. The fundamental property of the new vacuum is the restoration of thermal T-duality symmetry, implying a maximal critical temperature $T_c$, which occurs at the self-dual point. The duality invariant temperature can be written as $T=T_c\, e^{-\abs \sigma\abs}$, where $\sigma$ is the thermal modulus parametrizing the radius of the Euclidean time circle, $R_0=R_c\, e^{\sigma}$. It turns out that 
the right-moving sector gives rise to an {\it asymptotically supersymmetric structure} \cite{KutSeiberg,Misaligned} that restores  
thermal T-duality symmetry, avoiding at the same time the Hagedorn tachyonic instabilities \cite{dNonSingular}.

In all such thermal stringy systems, there are three characteristic regimes, each with a distinct effective field theory description: Two dual asymptotically cold regimes associated with the light thermal momentum and light thermal winding states respectively, and the intermediate regime where additional massless thermal states appear, leading to enhanced Euclidean gauge symmetry. The extra massless states source spacelike branes, localized at the critical points, which glue together the Momentum and Winding regimes. Thus the intermediate phase comprises a  ``Brane regime". As shown in \cite{FKPT,dNonSingular}, the thermal partition function exhibits a conical structure as a function of the thermal modulus $\sigma$, irrespectively of the dimensionality of the model. Thanks to the {\it asymptotically right-moving supersymmetric structure}, in both the Momentum $\{{\cal M}(\sigma> 0)\}$ and Winding $\{{\cal W}(\sigma<0)\}$ regimes, the energy density
and pressure can be well-approximated by the energy density and pressure of massless thermal 
radiation up to the critical temperature \cite{dNonSingular}:
\be
\rho\simeq (d-1) P,~~~~~~~P\simeq \,n^* \,\Sigma_d \, T^d\, ,~~~~~~~T=T_c ~e^{-\abs \sigma\abs} \,.
\ee
Here, $\Sigma_d$ is the Stefan-Boltzmann constant and $n^*$ is the number of initially massless states, modulo a spin-statistics factor for fermions:
$$
\Sigma_d={\Gamma(d/2)\over \pi^{d/2}}\, \zeta (d)~, ~~~~~~~n^* = n^B + n^F\, \left( {2^{d-1}-1 \over 2^{d-1}}\right).
$$
 The conical singularity in $\sigma $ is resolved by {\it the spacelike branes}, which comprise the ``Brane
 regime" $\{{\cal B}(\sigma=0)\}$ at the critical point  $\sigma=0$. These branes provide 
 localized (in time) negative pressure contributions, which turn out to be crucial in evading the constraints 
imposed by the singularity theorems of classical general relativity \cite{HE}
on realizing singularity-free, bouncing cosmologies.

 Utilizing the ingredients above, a string effective Lorentzian action covering simultaneously the three characteristic regimes was obtained in Refs \cite{FKPT,dNonSingular}. This action incorporates the spacelike branes that glue together: \\ 
 i) the {\it Winding regime $\equiv \{{\cal W}(\sigma<0;~ \tau<\tau_c)\}$}, \\
 ii) the {\it Brane regime $\equiv \{{\cal B}(\sigma=0; ~\tau=\tau_c)\}$} and \\
 iii)  the {\it Momentum regime  $\equiv \{{\cal M}(\sigma>0;~ \tau>\tau_c)\}$}\\
  at a given time $\tau_c $. The non-singular string cosmology can be viewed as the gluing at $\tau_c$ of the above three mentioned regimes:
\be 
\{{\cal C}_{\rm String}(\tau)\} \equiv  \{{\cal W}(\tau<\tau_c)\}  \oplus \{{\cal B}(\tau=\tau_c)\} \oplus \{{\cal M} (\tau>\tau_c)\}.
\ee
We would like to stress here that the existence of string mechanisms gluing distinct effective field theories were conjectured in the past by several authors and in several related contexts, like for instance the gluing of the $\N=2$ Calabi-Yau theories with the $\N=2$ Landau-Ginsbourg theories \cite{Witten:1993yc}, or even the gluing of theories related by S and/or T-dualities \cite{Dualities, CosmoTopologyChange}. In the majority of examples in the literature, the precise gluing mechanism could not be well established, since the intermediate region corresponded to a non-perturbative regime, and therefore 
the description was technically uncontrollable. 

The aim of this work is to investigate the possibility of constructing new interesting $d$-dimensional string cosmologies, 
 $\{{\cal \tilde C}_{String}(\tau) \}$, obtained by gluing at $\tau_c$ a Brane regime $\{\B(\tau\le \tau_c)\}$, which lasts arbitrarily long in time, with the Momentum $\{{\cal M}(\tau>\tau_c)\}$ regime: 
\be
\{ {\cal \tilde C}_{String}(\tau)\}\equiv \{ {\cal B}(\tau\le \tau_c)\} \oplus \{ {\cal M} (\tau>\tau_c)\}.
\ee
Thus, initially the Universe is described by the Brane regime. During this phase, the $\sigma$-model temperature and scale factor remain constant, with the temperature maintained at its maximal critical value $T_c$, while the string coupling grows from super-weak values in the very early past, reaching a maximal value $g^*_{\rm str}$ at $\tau_c$. Just after this moment, the Universe exits into the radiation dominated Momentum regime. 
In the Einstein frame, the whole evolution describes a bouncing Universe, with two distinct phases connected at $\tau=\tau_c$. In the initial contracting phase associated to the Brane regime, the scale factor evolves as in a negatively curved Universe filled with radiation. 
In the Momentum regime, the Universe is expanding with thermal radiation and a running dilaton. 
As we will show, the Brane regime can be understood in terms of a temporal distribution of the thin spacelike branes which appear in the cosmologies of \cite{FKPT,dNonSingular} at the critical point $\sigma=0$. The microscopic origin of the effective action in the Brane regime follows from the underlying stringy description of the thermal system at the extended symmetry point $\sigma=0$, incorporating various fluxes in the effective gauged supergravity theory.  The whole of the cosmological evolution can be treated perturbatively provided that the critical value of the string coupling $g^*_{\rm str}$ is small enough.

The plan of the paper is as follows.
In Section \ref{review}, we briefly review some of the type II thermal vacua, which are free of Hagedorn instabilities due to the presence of non-trivial gravito-magnetic fluxes. Some of their characteristic properties are displayed, such as the universal conical structure of the thermal partition function, the existence of a maximal critical temperature $T_c$, as well as the existence of the three distinct effective field theory descriptions associated with the Winding regime $\{ {\cal W}(\sigma<0)\}$, the Brane regime $\{{\cal B}(\sigma=0)\}$ and the Momentum regime $\{{\cal M}(\sigma>0)\}$. 
In Section \ref{EffectiveActions}, we present the effective actions valid in each of the three phases respectively. Special attention is devoted to the Brane regime. The origin of the latter is purely stringy, characterized by an enhancement of the Euclidean-time $U(1)_L$ gauge symmetry  
to an $[SU(2)_L]_{k=2}$ symmetry at $T_c$. At this critical temperature, additional massless thermal states source localized negative pressure contributions to the effective action, admitting a spacelike brane interpretation.  We explore the possibility of distributing the branes in the time interval $\tau\le \tau_c$. The consistency of such a distribution as well as the constraints imposed on the general structure of the Brane effective potential are analyzed in Section \ref{BraneDistributionV}.  
The distribution of branes gives rise to an interesting cosmology with the Universe being initially in the Brane regime $\{ \B(\tau\le \tau_c)\}$ up to $\tau_c$. This early times cosmological phase, as well as the Winding and Momentum cosmological phases are described in Sections \ref{Bcosmo} and \ref{Mcosmo}. 
In Section \ref{BtoMgluing}, the gluing of the initial Brane regime $\{ \B(\tau\le \tau_c)\}$ with the Momentum regime $\{ {\cal M}(\tau> \tau_c)\}$ is presented. The resulting string cosmologies $\{ {\cal \tilde C}_{\rm String}(\tau)\}\equiv \{ {\cal B}(\tau\le \tau_c)\} \oplus \{ {\cal M}(\tau>\tau_c)\}$, described 
in Section \ref{BMcosmo}, are free of initial curvature singularities and are compatible with string perturbation theory throughout the evolution. In Section \ref{Micro}, we present the microscopic origin of the Brane effective action, and in particular the Brane effective potential, in terms of non-trivial fluxes, making use of the underlying gauged supergravity effective description of the stringy extended symmetry point. Finally, our results and perspectives are summarized in Section \ref{conclu}.

\section{Modified thermal ensemble with fluxes, temperature duality and the Hagedorn transition }
\label{review}

In this Section, we review the key properties of tachyon-free thermal configurations associated to
type II $\N_4=(4,0)$ vacua in various dimensions. At zero temperature, 
the left-moving worldsheet degrees of freedom 
give rise to $16$ real supercharges, while the remaining right-moving supersymmetries 
are broken spontaneously via asymmetric geometrical fluxes \cite{akpt,massivesusy,FKT,FKPT,dNonSingular}. At special values of the moduli participating in the breaking of the right-moving supersymmetries, the local gauge symmetry is enhanced to a non-Abelian gauge symmetry. Thermal and quantum effects can stabilize 
the system at such extended symmetry points \cite{min,stabmod,CriticalCosmo}.
  
Finite temperature is introduced along with non-trivial gravito-magnetic fluxes, 
threading the Euclidean time cycle together with other cycles responsible for the breaking of the right-moving supersymmetries \cite{akpt,massivesusy,FKT,FKPT,dNonSingular}. 
These fluxes inject into the thermal vacuum non-trivial momentum and winding charges and lift the Hagedorn instabilities of the canonical ensemble, leading to a restoration of thermal duality symmetry 
for the partition function:  $Z(\beta / \beta_c)= Z(\beta_c/\beta)$. 
Here $\beta$ denotes the period of the Euclidean time cycle, 
attaining a critical value $\beta_c$ at the self-dual point. 
In all these models the absence of the conventional exponential growth 
of thermally excited massive states is due to the presence of asymptotic supersymmetry 
coming from the right-moving sector.
Although our explicit examples of tachyon-free thermal superstring models, which are described in great detail in the literature, 
involve type II initially $(4,0)$ vacua, we believe that the key properties, as identified 
in these models, will be generic in all thermal, tachyon-free superstring realizations\footnote{Similar behavior is exhibited in the two-dimensional
heterotic strings \cite{DLS}.}.  
 
The main properties which lead to the resolution of the Hagedorn and initial singularities are exhibited below: 
\begin{itemize}
 \item
The gravito-magnetic fluxes render the spectrum of thermal masses semi-positive definite. 
Consequently, the partition function is finite for all values of $\beta$ and duality invariant under $\beta \to \beta_c^2/\beta$. 
At the critical point $\beta=\beta_c$, 
new massless thermal states appear, extending the Euclidean-time $U(1)_L$ gauge symmetry 
to a non-Abelian $[SU(2)_L]_{k=2}$ symmetry. 
This phenomenon is absent in any conventional field theory model. 
In all string models, the self-dual point $\beta=\beta_c$ occurs at the fermionic point. 
\item
The extra massless states at the critical point possess non-trivial 
momentum and winding charges along the Euclidean-time circle, so that $p_L = \pm 1$ and $p_R=0$. 
These two extra states together with the thermal radius modulus 
give rise to the $SU(2)$ enhanced symmetry. At the critical point, 
the massless states give rise to non-trivial backgrounds,  
which admit a localized brane interpretation (in the Euclidean).   
\item
For ${\beta/\beta_c }\gg 1$, the thermal partition function 
is dominated by the light thermal momentum states, giving rise to
the characteristic behavior of massless thermal radiation in $d$ dimensions, 
modulo exponentially suppressed contributions from the massive string oscillator states:
\be
{Z \over V_{d-1}} ={n^* \Sigma_d\over \beta_c^{d-1}}\left({\beta_c \over \beta}\right)^{d-1}
+\;\O\left ( e ^{-\beta/\beta_c}  \right),
\ee 
where $n^*$ counts the number of effectively massless degrees of freedom, 
$\Sigma_d$ is the Stefan-Boltzmann constant for radiation and $V_{d-1}$ is the spatial volume. 
Thanks to the thermal duality symmetry, the asymptotic behavior for 
${\beta/\beta_c }\ll 1$ is dual-to-thermal, dominated by the light thermal winding states:
\be
{Z \over V_{d-1}} ={n^* \Sigma_d\over \beta_c^{d-1}}\left({\beta \over \beta_c}\right)^{d-1}
+\; \O\left ( e ^{-\beta_c/\beta}  \right).
\ee 
Here also, the oscillator states give exponentially suppressed contributions. 
The contribution of the massive oscillator states remains finite at the critical point, 
as the fluxes effectively reduce the density of thermally excited massive states. 
In most cases, the contribution of the massive oscillator states 
never dominates over the thermally excited massless states, due to asymptotic supersymmetry coming 
from the right-moving sector \cite{KutSeiberg,Misaligned,akpt,massivesusy,FKT,FKPT,dNonSingular}.  
\item
{\it The duality invariant temperature} $T$, 
valid in both asymptotic thermal regimes, is given by  $T\equiv T_c\,  e^{-\abs \sigma\abs}$, where 
we have defined the thermal modulus $\sigma$ 
by $ e^{\sigma}=\beta/\beta_c$. 
As a result, the temperature, the energy density and pressure in these configurations never exceed certain maximal values. 
The maximal critical temperature is given by $T_c = 1 / \beta_c$. In both asymptotic regimes
($T\ll T_c$), the partition function can be expressed in terms of the self-dual temperature as follows:
\be
{Z \over V_{d-1}} =n^* \,\Sigma_d\,T_c^{d-1}\left({T \over T_c}\right)^{d-1}
+\; \O\left ( e ^{-T_c/T}  \right).
\ee 
\item
Due to the right-moving asymptotic supersymmetry, the behavior of the thermal partition function $Z(\sigma)$ is controlled by 
the thermally excited massless states everywhere and up to the critical point $\sigma=0$
\cite{dNonSingular}:
\be
\label{local beha}
{Z \over V_{d-1}} \simeq n^*\Sigma_d\, T_c^{d-1}\,e^{-(d-1)\abs \sigma \abs},
\ee
modulo the exponentially suppressed contributions in all aforementioned regimes. 
This result reveals a universal conical structure as a function of the thermal modulus $\sigma$, irrespectively of the dimensionality of the model. 
The conical structure at the critical point is naturally resolved by the presence of additional massless thermal states.
This result also implies that in each of the two thermal phases corresponding to the light thermal momenta and light windings, 
the various thermodynamical quantities enjoy the standard monotonicity properties as functions of the temperature, with the
specific heat being positive up to the critical point.  
\item
The presence of the localized massless states is crucial since {\it they 
can marginally induce transitions between purely momentum and purely winding states}, 
thus driving a phase transition between the two dual asymptotically cold regimes. 
As in \cite{FKPT,dNonSingular}, this phase transition admits an almost geometrical description, 
in terms of a ``T-fold'' \cite{Hull:2006va,KTT}, with branes localized at the critical point gluing 
the ``Momentum'' and ``Winding'' spaces. The branes provide localized (in time) negative pressure contributions to the effective action. 
\end{itemize}

We conclude that the stringy thermal system has three characteristic regimes:
The regime of light thermal windings,  {\it $\{{\cal W}(\sigma<0)\}$}, 
the dual regime of light thermal momenta, {\it $\{{\cal M}(\sigma>0)\}$},  and the third intermediate brane regime, {\it $\{{\cal B}(\sigma=0)\}$}, corresponding to the extended symmetry point, where
additional massless thermal states carrying non-trivial momentum 
and winding charges become relevant.  In Refs \cite{FKPT,dNonSingular}, it was shown how to realize non-singular string cosmologies via a stringy gluing mechanism of the three regimes, which incorporates spacelike branes at a certain time $\tau_c$ when the temperature reaches its critical value:  
\be 
\{{\cal C}_{\rm String}(\tau)\} \equiv  \{{\cal W}(\tau<\tau_c)\}  \oplus \{{\cal B}(\tau=\tau_c)\} \oplus \{{\cal M} (\tau>\tau_c)\}.
\ee
The ingredients described above not only treat successfully the Hagedorn transition, 
but lead also to non-singular thermal cosmologies in contrast to field theoretic cases. 

The advantage of the above described cosmological examples is that the stringy gluing mechanism turns out to be under perturbative control, for all cosmological times, both under the string coupling $g_{\rm str}$ and the $\alpha^{\prime}$-expansions. Indeed, the Brane regime turns out to be described by an exact conformal field theory, which can be treated beyond any $\alpha^{\prime}$ approximation. Furthermore, the string coupling turns out to be bounded by a critical coupling $g^*_{\rm str}=g_{\rm str}(\tau_c)$, which can be chosen to be sufficiently small so that both 
the localized negative pressure contribution from the Brane regime as well as the bulk thermal corrections arising from the Winding and Momentum regimes can be determined unambiguously. These facts allow us to realize the gluing stringy mechanism 
and to obtain the non-singular cosmological evolution $\{ {\cal C}_{\rm String}(\tau) \}$ valid at all times. The resulting non-singular string cosmologies $\{ {\cal C}_{\rm String}(\tau)\}$ describe bouncing Universes, which in the asymptotic regimes  $\abs \tau \abs \gg 1$ can be either radiation or curvature dominated \cite{dNonSingular}. The bounce occurs at the brane regime $\{{\cal B} (\tau_c)\}$, where a phase transition occurs between the Winding $\{{\cal W}(\tau<\tau_c)\}$ and the Momentum 
$\{{\cal M}(\tau>\tau_c)\}$ regimes. These bouncing cosmologies are the first higher dimensional examples, where both the Hagedorn singularity 
and the classical Big Bang singularity are successfully resolved, remaining perturbative throughout the evolution.  

In this work we explore a new class of $d$-dimensional cosmologies, 
 $\{{\cal \tilde C}_{\rm String}(\tau) \}$, obtained by gluing at $\tau_c$ a Brane regime $\{ B(\tau\le \tau_c)\}$ with the Momentum regime $\{{\cal M}(\tau>\tau_c)\}$: 
\be
\{ {\cal \tilde C}_{\rm String}(t)\}\equiv \{ {\cal B}(\tau\le \tau_c)\} \oplus \{ {\cal M} (\tau>\tau_c)\}\, .
\ee
We would like to examine under what circumstances the Universe will remain in the Brane regime at the early times $\tau\le \tau_c$, while for $\tau> \tau_c$ will enter in the Momentum regime described by thermal radiation.


\section{The three effective field theory actions up to genus-1}
\label{EffectiveActions}

The above discussion and the results of Refs \cite{FKPT,dNonSingular}, summarized in the previous Section, clearly show that 
the cosmological evolution depends crucially on the value of the thermal modulus $\sigma$. There are three possible 
effective field theory actions associated to distinct cosmological time intervals:  $i$)  the Winding-action $\S_W$, which is restricted 
to $\sigma(\tau)< 0$,   $ii$) the Brane-action $\S_B$, with $\sigma(\tau)=0$ and iii) the Momentum-action $\S_M$ with $\sigma(\tau)>0$.

During the cosmological evolution, the thermal modulus $\sigma$ acquires non-trivial time-dependence, $\sigma(\tau)$, and so the Universe 
may pass through all of the three regimes. The transition from the Winding to the Momentum regime (and vice versa), 
necessarily crosses the Brane regime, where the Universe can stay in principle for a certain amount of time. Therefore, in order to study the stringy 
cosmological evolution, it is necessary not only to derive the three effective actions, but also to specify the gluing mechanism of
the Winding with the Brane regime as well as the Brane with the Momentum regime.

The Momentum and Winding regimes can be easily described by effective $d$-dimensional dilaton-gravity $\sigma$-model actions (up to the genus-1 level):
\begin{align}
&\S_M = \int d^d x\, \sqrt{-g }\, \Theta (\sigma)\left\{e^{-2\phi}\,\left[\frac{\cal R}{2}+2(\partial\phi)^2 +\cdots \right]  + P(\sigma ) \right\}\desp\nonumber\\
\label{SMW}
&\S_W= \int d^d x\, \sqrt{-g }\, \Theta (-\sigma)\left\{e^{-2\phi}\,\left[\frac{\cal R}{2}+2(\partial\phi)^2 +\cdots \right]  + P(-\sigma ) \right\}~.
\end{align}
 The term proportional to the pressure $P$ in both expressions is the genus-1 contribution of the thermal effective potential, 
while the rest comprise the genus-0 effective dilaton-gravity action written in the $\sigma$-model frame.
The ellipses denote the contributions of the two-index antisymmetric tensor, other moduli fields, gauge bosons and space-time fermions. 
The two actions, modulo the $\Theta$-constraints, look identical. 
This is a consequence of thermal duality, which implies that the pressure 
\be
P\simeq n^* \Sigma_d\,T_c^d \, e^{-d\abs \sigma \abs}\,
\ee
is a duality invariant quantity, even though the spinor representations in the Winding regime have opposite chirality from those in the 
Momentum regime \cite{akpt,FKPT,dNonSingular}. The Spinor and anti-Spinor contributions to the thermal effective potential amount to identical results. 
Moreover, all thermodynamical quantities such as the temperature, the energy density and the pressure are given in terms of manifestly 
duality invariant expressions involving the absolute value of the thermal modulus $\abs \sigma \abs$. Thus thanks to thermal duality, both the Momentum 
and Winding regimes can be simultaneously described by {\it a unique expansion in terms of the duality invariant temperature} 
$T= T_c\, e^{-\abs \sigma \abs}$. 

The Brane regime appears when $\sigma=0$. Its origin is purely stringy, characterized in the Euclidean by extra massless 
thermal states carrying non-trivial winding and momentum charges. The existence of such states is crucial for realizing the gluing 
mechanism between the Winding, Brane and Momentum phases. It is also interesting that the Brane regime is well described in terms of an $[SU(2)_L]_{k=2}$ 
conformal field theory associated with the fermionic extended symmetry point. In this regime, we have to include the genus-0 backgrounds 
of the extra massless thermal states. 
These tree level contributions admit a brane interpretation \cite{FKPT,dNonSingular}. The brane tension is determined by the allowed
non-trivial backgrounds of the extra massless thermal scalars $\langle\partial\varphi^I\rangle^2\ne 0$, where there are non-trivial gradients 
along the directions transverse to Euclidean time, as well as non-trivial fluxes $\langle F^a_{IJ}\rangle\ne 0 $ 
associated to the extra gauge bosons appearing at $\tau_i$ with $\sigma(\tau_i)=0$.  
The non-zero values of $\langle\partial\varphi^I\rangle$ and $\langle F^a_{IJ}\rangle$ give rise to an effective localized Brane-action with a non-trivial potential 
for the dilaton field:
\be
\S_{B}(\tau_i) = \int d\tau \delta (\tau-\tau_i)\,d^{d-1}x\, \sqrt{g_\bot} \left\{ e^{-2\phi}\left[{{\cal R}\over 2}+2(\partial\phi)^2\right]+ P-V_i (\phi)\right\} .
\ee
On shell, the thermal modulus is identically vanishing and  the genus-1 contribution to the pressure is $P_c:=P(\sigma=0)$.  
The microscopic origin of the above action  and the precise structure of $V_i(\phi)$ will be presented in Section \ref{Micro}. 
In what follows we treat $V_i(\phi)$ as a generic function of $\phi$, specified when necessary.

In Refs \cite{FKPT,dNonSingular} the possibility that the brane appears only once at $\tau_c$ (with $\sigma(\tau_c)=0$) was investigated. 
The resulting cosmology gives rise to a bouncing Universe gluing together at $\tau_c$ a ``Winding" contracting phase 
with a ``Momentum" expanding phase. 
However, the brane picture suggests for another interesting possibility; namely a continuous brane distribution 
in an interval of time: $\tau_- \le \tau \le \tau_+$ with $\sigma(\tau)=0$. In these circumstances, the effective Brane-action takes the following 
form:
\begin{align}
\S_{B} = & \!\int \!d\tau^{\prime} \abs g_{00} \abs d\tau \delta (\tau-\tau^{\prime}) \Theta(\tau^{\prime}-\tau_-) \Theta(\tau_+-\tau^{\prime})d^{d-1}x\sqrt{g_\bot} \left\{e^{-2\phi}\!\left[{{\cal R}\over 2}+2(\partial\phi)^2\right]\!+\! P-V(\phi)\right\}\nonumber\\
&+\S_-+\S_+.
\end{align}
Introducing the time interval $D\equiv(\tau_- ,\, \tau_+)$,  the Brane-action can be written as:
\be
\label{SD}
\S_B=\int_{D}  d^d x\,\sqrt{-g}\left\{e^{-2\phi}\left[{{\cal R}\over 2}+2(\partial\phi)^2\right] +P-V(\phi)\right\}+\S_-+\S_+,
\ee
where the localized brane terms at  $\tau_{\pm}$ are  
\be
\label{S+-}
\S_\pm=- \int d\tau \delta (\tau-\tau_{\pm})\,d^{d-1}x\,\sqrt{g_\bot} \, V_{\pm}(\phi).
\ee
As we will see later these terms turn out to be crucial for achieving the gluing between the Brane with the Momentum and/or the Winding regime(s).

The string effective action and the resulting cosmological solutions (defined for all times),
 can be obtained once we combine consistently the three regimes, namely:
\be
\S=\S_W+\S_B+\S_M,
\ee
so that the thermal modulus $\sigma$ is unrestricted in $\S$, 
even thought $\sigma$ is restricted in the individual effective actions $S_W,~S_M$ and $S_B$.

\subsection{Consistency conditions for $V(\phi)$ in the Brane regime }
\label{BraneDistributionV}
 
In the Brane-action, we have introduced an effective potential $V(\phi)$. We will see in Section \ref{Micro} that this potential is sourced by the extra massless states, 
which give rise to non-trivial gradients and fluxes at $\sigma=0$, \ie when the temperature reaches its maximal value $T=T_c$. Thus, during the Brane regime, 
the temperature must remain constant. This implies that the structure of the effective potential is not arbitrary but strongly 
restricted by the constancy of the temperature $T$ during the time interval $D=(\tau_- ,\tau_+)$.  
We are interested in homogeneous and isotropic backgrounds
\be
ds^2=-N(\tau)^2{d\tau}^2+a(\tau)^2d\Omega_k^2,
\ee
where $\Omega_k$ is a $(d-1)$-dimensional Einstein space with curvature $k\le 0$. 
The equations of motion for the laps function $N$  and the scale factor $a$ are ($H\equiv\dot a/a$) :
\begin{align}
\label{eqN}
&(N):\quad {1\over2}(d-1)(d-2) \left( H^2+ k{N^2\over a^2} \right) =2(d-1)H\dot \phi-2\dot\phi^2+e^{2\phi}N^2(\rho+V),\esp\\
& (a): \quad (d-2)\left({\ddot a\over a} - H{\dot N\over N}\right)+{1\over2}(d-2)(d-3)  \left( H^2+ k{N^2\over a^2} \right)=\nonumber\\
&\label{eqab}\qquad \qquad\qquad\qquad\qquad\qquad  2\ddot\phi+2(d-2)H\dot\phi-2\dot\phi^2-2{\dot N\over N}\dot\phi-e^{2\phi}N^2(P-V).
\end{align}
The dilaton equation is given by
\be
\label{eqphi}
\ddot\phi+(d-1)H\dot\phi-\dot\phi^2-{\dot N\over N}\dot\phi-{d-1\over 2}\left({\ddot a\over a} - H{\dot N\over N}\right)
-{1\over 4}(d-1)(d-2)\left( H^2+ k{N^2\over a^2} \right)={1\over 4}e^{2\phi}N^2\, {dV\over d\phi}. 
\ee
Combining the above equations and using the thermodynamical identity $\rho=-P-{\partial  \over \partial \abs \sigma\abs} P$, 
we obtain the entropy conservation equation,
\be 
(\dot \rho+\dot P)+ \big( (d-1)H +\abs \dot \sigma\abs \big) (\rho+P)=0,
\ee
which implies that in order for the temperature to be constant, attaining its maximal value $T=T_c$ (equivalently $\sigma=0)$, 
not only are the pressure and the energy density constant, ($\rho=\rho_c,\, P=P_c$), 
but also the scale factor must be constant: $a=a_c$ with $H=0$. 
Then, in the conformal gauge $N\equiv a$, where $a\equiv a_c$, the equations of motion yield:
\begin{align}
\label{eqN'}
&(N): ~~{1\over2}(d-1)(d-2) k =-2\dot\phi^2+e^{2\phi}a_c^2(\rho_c+V),\esp\\
\label{eqa'}
&(a):~~{1\over2}(d-2)(d-3)  k=2\ddot\phi-2\dot\phi^2-e^{2\phi}a_c^2(P_c-V),\esp\\
\label{eqphi'}
&(\phi):~~\ddot\phi-\dot\phi^2-{1\over 4}(d-1)(d-2) k={1\over 4}e^{2\phi}a_c^2\, {dV\over d\phi}. 
\end{align}
Combining Eqs $(a)$  and $(\phi)$, we obtain an equation constraining the structure of the effective potential $V(\phi)$:
\be
\label{eqV}
{dV\over d\phi}+2V=2P_c-2(d-2){k\over a_c^2}\, e^{-2\phi}.
\ee
This can be integrated to find:
\be
V(\phi)=C\, e^{-2\phi}+B-2(d-2){k\over a_c^2}\phi \, e^{-2\phi}\qquad\where \qquad B=P_c
\ee
and $C$ is an arbitrary integration constant. Eqs $(N)$ and $(a)$ then become:
\begin{align}
\label{eqN''}
&{1\over2}(d-1)(d-2) k =-2\dot\phi^2+a_c^2\big[(\rho_c+P_c)e^{2\phi}+C]-2(d-2)k\phi,\esp\\
\label{eqa''}
&{1\over2}(d-2)(d-3)  k=2\ddot\phi-2\dot\phi^2+a_c^2C-2(d-2)k\phi.
\end{align}
In the following, only Eq. (\ref{eqN''}) needs  to be solve, since it is easily seen to imply Eq. (\ref{eqa''}). Restricting to the case $k=0$, the structure of the effective potential has a very suggestive form.
Namely, it is given in terms of a binomial of $e^{-2\phi}$. Its origin has a well defined interpretation in terms of internal graviphoton 
and matter gauge field fluxes. We will return to this point in Section \ref{Micro}. 

\subsection{Cosmological evolution in the Brane regime, for $k=0$}
\label{Bcosmo}

In the $\sigma$-model frame, the temperature $T$ and the scale factor $a$ are fixed to their critical values during the Brane regime. 
The only non-trivial evolution is that of the dilaton field, which can be easily derived in terms of the flux parameters $C$ and $B$ 
appearing in the effective potential. Actually, there is only one free parameter, since $B$ is fixed to the critical value of the pressure, $B=P_c$.  

In what follows, we restrict to the $k=0$ case. When $C\ge 0$,  we set the earlier endpoint of the Brane regime at infinity, $\tau_-=-\infty$. This means that for $\tau\le \tau_+$, the Universe is in the Brane regime with a flat $\sigma$-model metric and 
constant critical temperature. The evolution of the dilaton as a function of conformal time ($N\equiv a$) is found by solving Eq. (\ref{eqN''}). When $C=0$, one finds
\be
\label{C=0}
C=0\; : \qquad e^{-\phi}=a_c\,  \sqrt{\rho_c+P_c\over 2}\, (- \tau),\qquad a=a_c, \qquad T=T_c,\qquad \forall \tau\le \tau_+<0.
\ee
The upper bound of $\tau$ is chosen to be strictly negative, $\tau_+<0$, in order to maintain the validity of the perturbative approach in the Brane regime. The choice of $\tau_+$ determines the string coupling at the transition towards the Momentum phase, $\phi_+:=\phi(\tau_+)$. 
For $C>0$, the solution takes the form:
\be
\label{C>0}
C>0\; : \quad e^{-\phi}=\sqrt{\rho_c+P_c\over C}\,\sinh\Big[a_c\sqrt{C\over 2}(- \tau)\Big],\quad a=a_c, \quad T=T_c,\quad \forall \tau\le \tau_+<0,
\ee
where the upper bound $\tau_+<0$ guaranties the validity of the perturbative approach as before. For $C<0$, the evolution of the dilaton becomes 
\be
\label{C<0}
 C<0\; : \quad e^{-\phi}=\sqrt{\rho_c+P_c\over -C} \,\sin\Big[a_c\sqrt{-C\over 2}(- \tau)\Big],\quad a=a_c, \quad T=T_c,\quad \forall \tau_-\le \tau\le \tau_+<0.
\ee
In this case, in order to avoid  a second non-perturbative regime, we need to impose a lower bound for $\tau$ which satisfies 
\be
\label{taumin}
\tau_{\rm min}:=-{\pi\over a_c}\, \sqrt{2\over -C}<\tau_-.
\ee
Thus, the Brane regime lasts for a suitable finite time interval in this case. The choice of $\tau_-$ determines the string coupling when the Universe enters the Brane regime from the Winding phase, while $\tau_+$ controls the dilaton at the exit of the Brane regime towards the Momentum phase, $\phi_\pm:=\phi(\tau_\pm)$.  

Although the metric is flat in the $\sigma$-model frame, it acquires a non-trivial time-dependence in the Einstein frame defined for $d>2$, 
due to the rescaling by a dilaton-dependent factor: 
${g}^E_{\mu\nu}\equiv e^{-{4\phi\over d-2}}\, g_{\mu\nu}$. Consequently, the laps function, scale factor and temperature in the Einstein frame are given by
\be
N_E=N\,e^{-{2\phi\over d-2}}\; ,\qquad   a_E=a_c\,e^{-{2\phi\over d-2}}\; , \qquad T_E=T_c\,e^{{2\phi\over d-2}}.
\ee
To determine them in the cosmological frame defined as
\be
\label{cosmoframe}
ds^2_E=-dt^2+{a}_E^2(t)\, d\v{\mathbf x}^2,
\ee
we rewrite Eq. (\ref{eqN''}) in the gauge $N\equiv e^{{2\phi\over d-2}}$:
\be
\label{Fr}
2\dot\phi^2=e^{{4\phi\over d-2}}\, \big[(\rho_c+P_c)\, e^{2\phi}+C\big].
\ee
To interpret the above equation, we express it in terms of the scale factor in the Einstein frame. Defining $H_E\equiv\dot a_E/a_E$, the dilaton kinetic energy on the left hand side is proportional to $H_E^2$, while on the right hand side, the terms with coefficients $(\rho_c+P_c)$ and $C$ are proportional to $1/a^d$ and $1/a^2$, respectively. To be specific, one obtains
\be
\label{Frieeff}
{1\over 2}(d-1)(d-2)\left(H_E^2+{k_{\rm eff}\over a_E^2}\right)={r_{\rm eff}\over  a_E^d}\, ,
\ee
where 
\be
k_{\rm eff}=-C\, {2a_c^2\over (d-2)^2}\qquad \and\qquad r_{\rm eff}={d-1\over d-2}\, (\rho_c+P_c)\, a_c^d.
\ee
This shows that effectively, the evolution of the Universe is identical to that of a spatially curved space filled with thermal radiation. In fact, denoting by $n^*_{\rm eff}$ the effective number of massless degrees of freedom associated to this radiation-like energy density, 
\be
{r_{\rm eff}\over a_c^d}:=(d-1)\, {n^*_{\rm eff}\, \Sigma_d\over \beta_c^d},
\ee
we find $n^*_{\rm eff}$ is larger than $n^*$,
\be
n^*_{\rm eff}={d\over d-2}\, n^*.
\ee

When $C=0$, the thermal energy $\rho_c$ and the flux contribution $B=P_c$ yield a radiation-like evolution
\be
\label{C=0a}
C=0\; : \quad e^{-{d\phi\over d-2}}=\left({T_c\over T_E}\right)^{d\over 2}=\left({a_E\over a_c}\right)^{d\over 2}={d\over d-2}\sqrt{\rho_c+P_c\over 2}\, (-t) , \quad \forall t\le t_+<0,
\ee
where the upper bound $t_+$ of the cosmological time is chosen negative, in order to avoid a non-perturbative regime and a formal Big Crunch at $t=0$.  To find the evolution when $C\neq 0$, we define $s=\sign(C)$ and
\be
\label{v}
v= \left({\abs  C\abs\over \rho_c+P_c}\right)^{d\over 2(d-2)}\,  \left({a_E\over a_c}\right)^{d\over 2},
\ee
to rewrite the effective Friedmann equation (\ref{Frieeff}) as
\be
{dv\over \sqrt{1+s\, v^{2(d-2)\over d}}}=\pm \Omega\, dt\qquad \where\qquad \Omega={d\over d-2}\sqrt{\rho_c+P_c\over 2}\left({\abs C\abs\over \rho_c+P_c}\right)^{d\over 2(d-2)}.
\ee
Defining the primitive 
\be
f^s_d(v)=\int_0^v {dz\over\sqrt{1+s\, z^{2(d-2)\over d}}}\equiv v\; {}_2F_1\Big({1\over 2},{d\over 2(d-2)};1+{d\over 2(d-2)}; -s\, v^{2(d-2)\over d}\Big),
\label{exp}
\ee
we obtain
\be
\label{eqv}
f^s_d(v)=-\Omega\,  t,\qquad \forall t<0.
\ee
The function $f^+_d(v)$ for $v>0$ is monotonically increasing and can be inverted. As a result, the  cosmological evolution for $C>0$ takes the form
\be
\label{C>0E}
C> 0\; : \quad e^{-{d\phi\over d-2}}=\left({T_c\over T_E}\right)^{d\over 2}=\left({a_E\over a_c}\right)^{d\over 2}=\left({\rho_c+P_c\over C}\right)^{d\over 2(d-2)}f_d^{+(-1)}(-\Omega \,t),\quad \forall t\le t_+<0.
\ee
The choice of an upper bound $t_+<0$ for the cosmological time guaranties that the Universe passes into the Momentum phase and avoids a strong coupling regime. 
The function $f^-_d(v)$ is defined for $0<v\le1$, where it is monotonically increasing and invertible. The time-evolution we find is
\be
\label{C<0E}
C< 0\; : \; e^{-{d\phi\over d-2}}=\left({T_c\over T_E}\right)^{d\over 2}=\left({a_E\over a_c}\right)^{d\over 2}=\left({\rho_c+P_c\over -C}\right)^{d\over 2(d-2)}f_d^{-(-1)}(-\Omega \,t),\; \forall t_-\le t\le t_+<0.
\ee
In order to avoid two non-perturbative regimes, time is restricted to a range $t\in (t_-,t_+)$. As before, $t_+$ must be strictly negative, while $t_-$ satisfies
\be
t_{\rm min}:=-{2\over \Omega}\, f^{-(-1)}_d(1)\equiv-{2\sqrt{\pi}\over \Omega}\, {\Gamma(1+{d\over 2(d-2)})\over \Gamma(1+{1\over d-2})}<t_-.
\ee  
In fact, the formal solution for $C<0$ and $t_{\rm min}<t<0$ admits a symmetry, $t\to t_{\rm min}-t$. The latter maps the formal non-perturbative Big Crunch at $t=0$ to a formal non-perturbative Big Bang at $t=t_{\rm min}$. Moreover, the scale factor bounces when it reaches its maximum at $t=t_{\rm min}/2$.
The qualitative behavior of the above solutions for arbitrary $C$ is independent of the dimension $d>2$. For instance, in the physically interesting case $d=4$, the  analytic functions $f_d^{s(-1)}$ happen to be polynomials of degree two and the solutions take the explicit form
\be
\forall C\in \R\; : \quad e^{-2\phi}=\left({T_c\over T_E}\right)^2=\left({a_E\over a_c}\right)^2=\sqrt{2(\rho_c+P_c)}\, (-t)+{C\over 2}\,t^2.
\ee

The picture in Einstein frame can be summarized as follows:
\begin{itemize}
\item For $C=0$, we get effectively a contracting ``radiation-like" era (${H}_E^2\propto 1/a_E^d$), which starts at weak coupling. No singular behavior is encountered at large negative times, consistently with the choice $t_-=-\infty$. The Big Crunch and strong coupling regime at $t=0$ is avoided since the endpoint of the Brane regime is located at $t_+<0$. 
\item For $C>0$, we start with a weakly coupled contracting ``curvature-like'' dominated era ($k_{\rm eff} <0$ and $H^2_E\sim 1/{a}_E^2$), followed by a radiation dominated era. Here also, the Big Crunch and strong coupling regime is avoided since $t_+<0$.
\item For $C<0$, the effective curvature is positive ($k_{\rm eff} >0$) and the mathematical solution   starts formally with a Big Bang at infinite coupling, bounces and ends in a formal Big Crunch at infinite coupling. However, restricting time to the interval $(t_-,t_+)$ guaranties the  Brane regime to be compatible with our perturbative approximation. Depending on the fact that $t_{\rm min}/2$ may be lower than $t_-$, in the interval $(t_-,t_+)$ or larger than $t_+$, the scale factor is either decreasing, bouncing or increasing. 
\item In the cases we are mostly interested, we have $C\ge 0$ and $t_-$ can be consistently taken to $-\infty$. No singularity is encountered since $t_+<0$. We exit from the Brane regime at $t=t_+$, when the dilaton reaches a sufficiently large value $\phi_+$ for the brane to decay into the conventional perturbative radiation plus moving dilaton  era.  
\end{itemize}

\subsection{Cosmological evolution in the Momentum or Winding regimes, for $k=0$}
\label{Mcosmo}

For $\tau>\tau_+$, the Brane potential $V(\phi)$ and the localized terms responsible for the gluing of the Brane and Momentum regimes are absent. The evolution of the scale factor $a$, 
the temperature $T$ and the dilaton field $\phi$ are dictated by the Momentum effective action $\S_M$ given in Eq. (\ref{SMW}).  
In Ref. \cite{dNonSingular}, the cosmological evolution in this regime was determined for $d\ge 2$. 
The scale factor and temperature in $\sigma$-model frame are expressed in terms of the positive thermal modulus $\sigma$, 
\be
\label{confor}
{a\over a_c}={T_c\over T}=e^\sigma.
\ee
Under the approximation of Eq. (\ref{local beha}), which implies the state equation for radiation $\rho\simeq (d-1)P$, the modulus $\sigma$ and the dilaton field can be found analytically. For $d>2$ and in conformal gauge ($N\equiv a$), they are given by\footnote{A solution for $d=2$ also exists \cite{FKPT}. It can be recovered by taking the limit $d\to 2_+$ in Eq. (\ref{sol d}) \cite{dNonSingular}. In the Hybrid model, Eq. (\ref{local beha}) with $d=2$ is exact.}
\begin{align}
\label{sol d}
& \dis \sigma(\tau) ={1\over d-2}~\left[ {\eta_+} \ln \left(1+ {\omega_+ (\tau-\tau_+) \over \eta_+} \right) -{\eta_-}  \ln \left(1+{\omega_+ (\tau-\tau_+)\over \eta_-}  \right) \right],\quad \forall \tau>\tau_+, \desp\nonumber \\
&\dis \phi(\tau)=\phi_++{\sqrt{d-1}\over 2}~\left[ \ln \left(1+ {\omega_+ (\tau-\tau_+)\over \eta_+} \right) - \ln \left(1+ {\omega_+ (\tau-\tau_+)\over \eta_-} \right)  \right],
\end{align}
where  $\eta_\pm=\sqrt{d-1}\pm 1$. The parameter $\omega_+$ turns out to be related 
to the thermal energy density $\rho$, the scale factor and dilaton field at the instant $\tau_+$, when the Brane and Momentum regimes are glued:
\be
\label{omegakappaA}
\omega_+ = {d-2\over \sqrt{2(d-1)}} \, a_c\, \sqrt{\rho_c}\, e^{\phi_+}. 
\ee
In the Brane regime, the $\sigma$-model metric is static and the Ricci scalar  vanishes. In the Momentum regime, the Ricci scalar is decreasing with time and its maximal value is reached at the brane exit, 
\be
{\cal R}_+=2\rho_c\, e^{2\phi_+}.
\ee
Thus, both higher derivative corrections and higher genus contributions remain small throughout the Momentum regime and can be consistently neglected, provided the critical value of the string coupling $e^{\phi_+}$  is taken sufficiently small. 
Far from the brane ($\omega_+ (\tau-\tau_+) \gg 1$), the dilaton is asymptotically constant, the temperature drops and the scale factor tends to infinity. 

When the Brane regime is characterized by a potential $V(\phi)$ with a parameter $C<0$, it  must last a finite amount of time $(\tau_-,\tau_+)$ preceded by a Winding phase. For $\tau<\tau_-$, the thermal modulus is negative,
\be
\label{confor-}
{a\over a_c}={T_c\over T}=e^{-\sigma}.
\ee
It follows together with the dilaton field a time-evolution similar to that of the Momentum phase, 
\begin{align}
\label{sol d-}
& \dis \sigma(\tau) ={-1\over d-2}~\left[ {\eta_+} \ln \left(1+ {\omega_- (\tau_--\tau) \over \eta_+} \right) -{\eta_-}  \ln \left(1+{\omega_-  (\tau_--\tau)\over \eta_-}  \right) \right],\quad \forall \tau<\tau_-, \desp\nonumber \\
&\dis \phi(\tau)=\phi_-+{\sqrt{d-1}\over 2}~\left[ \ln \left(1+ {\omega_- (\tau_--\tau)\over \eta_+} \right) - \ln \left(1+ {\omega_- (\tau_--\tau)\over \eta_-} \right)  \right],
\end{align}
 where 
 \be
\label{omegakappa}
\omega_- = {d-2\over \sqrt{2(d-1)}} \, a_c\, \sqrt{\rho_c}\, e^{\phi_-}. 
\ee

The Momentum regime and the eventual Winding regime evolutions can be described in cosmological frame defined in Eq. (\ref{cosmoframe}) by solving the associated Friedmann equation. The latter takes the form
\be
\label{FrieMW}
{1\over 2}(d-1)(d-2)\,H_E^2={r_\pm\over  a_E^d}+{c_\pm\over a_E^{2(d-1)}}\, ,
\ee
where 
\be
r_\pm=\rho_c\, a_c^d\qquad \and\qquad c_\pm={1\over d-2}\, \rho_c\, e^{-2\phi_\pm}\, a_c^{2(d-1)}.
\ee
In the right hand side, the first contribution arises from the thermal radiation, while the second is the kinetic energy density of the dilaton modulus. Clearly, the late time evolution of the Universe ($t\gg \abs t_+\abs$) and the eventual early time one ($ t \ll t_-$)  are radiation dominated, while the contribution of the energy stored in the dilaton motion increases as we approach the transitions to the Brane regime, ($t-t_+\to 0_+$, $t-t_-\to 0_-$). 


\subsection{Gluing the Brane, Momentum and Winding regimes}
 \label{BtoMgluing}
Assuming $\tau_-=-\infty$, we first consider the string effective action which is valid in both the Brane and the Momentum regimes,
\be
\S_{BM}=\!\int \!d^d x\sqrt{-g}\left\{e^{-2\phi}\left[{{\cal R}\over 2}+2(\partial\phi)^2\right]\!+P-V(\phi)\Theta(\tau_+-\tau)\right\}-\!\int \!d^dx \sqrt{g_\bot}\, e^{-2\phi}\, \kappa_+\, \delta(\tau-\tau_+).
\ee
In this expression, we use the potential defined in Eq. (\ref{S+-}), which is valid at the endpoint $\tau_+$ of the brane regime: $V_+(\phi)=\kappa_+e^{-2\phi}$. The latter is derived in Refs \cite{FKPT,dNonSingular} and reviewed in Section \ref{kappa}. On the contrary, the effective potential $V(\phi)$ in the Brane regime arises from the distribution of thin branes 
in the time-interval $-\infty<\tau<\tau_+$. 
The above action leads to the following equations of motion:

\noindent - For $N$:
\be
\label{eqNtot}
{1\over2}(d-1)(d-2) \left( H^2+ k{N^2\over a^2} \right) =2(d-1)H\dot \phi-2\dot\phi^2+e^{2\phi}N^2\big(\rho+V\Theta(\tau_+-\tau)\big).
\ee

\noindent - For $a$:
\begin{align}
& (d-2)\left({\ddot a\over a} - H{\dot N\over N}\right)+{1\over2}(d-2)(d-3)  \left( H^2+ k{N^2\over a^2} \right)=\nonumber\\
&\label{eqa}\qquad 2\ddot\phi+2(d-2)H\dot\phi-2\dot\phi^2-2{\dot N\over N}\dot\phi-e^{2\phi}N^2\big(P-V\Theta(\tau_+-\tau)\big)+\kappa_+ N\delta(\tau-\tau_+).
\end{align}

\noindent - For $\phi$:
\begin{align}
&\ddot\phi+(d-1)H\dot\phi-\dot\phi^2-{\dot N\over N}\dot\phi-{d-1\over 2}\left({\ddot a\over a} - H{\dot N\over N}\right)\desp\nonumber \\
\label{eqphitot}&\quad -{1\over 4}(d-1)(d-2)\left( H^2+ k{N^2\over a^2} \right)={1\over 4}e^{2\phi}N^2\, {dV\over d\phi}\Theta(\tau_+-\tau)-{1\over 2}\kappa_+ N\delta(\tau-\tau_+). 
\end{align}
The above differential system  implies that $\dot\phi$ is discontinuous across the endpoint $\tau_+$ of the Brane, while $\dot a$ is smooth. Since the scale factor remains constant during  the Brane regime, $a\equiv a_c $ for $\tau<\tau_+$, it follows that $\dot a(\tau_+)=0$ at the transition to the Momentum phase. Consistently, this condition is fulfilled by the evolution displayed in Eqs (\ref{confor}) and (\ref{sol d}). The brane tension $\kappa_+$ is related to the discontinuity of $\dot \phi$ at $\tau=\tau_+$,
\be
\kappa_+=2\, {\dot\phi(\tau_+)_B-\dot\phi(\tau_+)_M\over N(\tau_+)}\, ,
\ee
as follows from Eq. (\ref{eqphitot}). Utilizing the Friedmann equation (\ref{eqNtot})  just before and after the brane endpoint $\tau_+$, the fact that $\dot a(\tau_+)=0$  and restricting to the case $k=0$, we obtain for $C\ge 0$,
\be
\dot\phi(\tau_+)_B={N(\tau_+)\over \sqrt{2}}\, \sqrt{V(\phi_+)+\rho_c}\, e^{\phi_+}\, ,\qquad \dot\phi(\tau_+)_M=-{N(\tau_+)\over \sqrt{2}}\, \sqrt{\rho_c}\, e^{\phi_+}\, ,
\ee
in terms of $\rho_c $ and $V(\phi_+)=P_c+C\,e^{-2\phi_+}$, where $\phi_+$ is the value of the dilaton field at $\tau_+$. Combining the above expressions, 
we obtain the brane tension $\kappa_+$,
\be
C\ge0 :\quad \kappa_+=\sqrt{2}\left(\sqrt{C+(\rho_c+P_c)\, e^{2\phi_+}}+ \sqrt{\rho_c}\, e^{\phi_+}\right).
\ee
Thus, $\kappa_+$ is determined by the parameter $C$, modulo corrections in string coupling, $\kappa_+=\sqrt{2C}+\O(e^{\phi_+})$. Note that the gluing of Winding and Brane regimes at some earlier instant $\tau_-$ would be incompatible with the positivity condition of a tension $\kappa_-$ at the transition. This is the reason why we set $\tau_-=-\infty$ and consider a gluing  at $\tau_+$ only.  

On the contrary, as we already stated in  Section \ref{Bcosmo}, the case $C<0$ requires the Brane regime to be restricted to a finite range of time, in order to avoid two non-perturbative regimes. The relevant effective action is the combination of the Winding-, Brane- and Momentum-actions: $\S_{WBM}$. Reasoning as before, the brane tension $\kappa_+$ that triggers the Brane to Momentum phase transition is found to be
 \be
C<0:\quad \kappa_+=\sqrt{2}\left(\sqrt{\rho_c}\, e^{\phi_c}+\epsilon_+ \sqrt{(\rho_c+P_c)\, e^{2\phi_c}-\abs C\abs} \right), 
\ee
where 
\be
\epsilon_+=\sign\big(\tau_+-{\tau_{\rm min}\over 2}\big)\quad \and\quad \tau_{\rm min}+{1\over a_c}\sqrt{2\over -C}\arcsin\left(\sqrt{P_c\over \rho_c+P_c}\right)\le \tau_+<0.
\ee
As compared to Eqs (\ref{C<0}) and (\ref{taumin}), the stronger constraint on the allowed choices of upper bound $\tau_+$ of the Brane regime follows from the positivity constraint to be imposed on $\kappa_+$. For $\tau<\tau_-$, we are in the 
Winding regime describing a contracting Universe with increasing dilaton.  At the transition from the Winding to the Brane regimes occurring at $\tau=\tau_-$, we have a tension
\be
C<0:\quad \kappa_-=\sqrt{2}\left(\sqrt{\rho_-}\, e^{\phi_c}+\epsilon_- \sqrt{(\rho_c+P_c)\, e^{2\phi_-}-\abs C\abs} \right), 
\ee
where $\phi_-$ is the dilaton field at $\tau_-$,
\be
\epsilon_-=\sign\big({\tau_{\rm min}\over 2}-\tau_-\big)\quad \and\quad \tau_{\rm min}\le \tau_-<-{1\over a_c}\sqrt{2\over -C}\arcsin\left(\sqrt{P_c\over \rho_c+P_c}\right).
\ee
Again, the positivity condition on $\kappa_-$ reduces the upper bound of the allowed values of the instant $\tau_-$  of entry  into the Brane regime. 
 
\subsection{String non-singular cosmologies}
 \label{BMcosmo}
 
 As we have shown in the previous Sections and in Refs \cite{FKPT,dNonSingular}, string theory provides the tools for realizing a gluing mechanism between different effective field theories valid in different cosmological regimes. This stringy gluing mechanism is possible due to the appearance of extra Euclidean massless states, localized at times when the temperature reaches its maximal critical value, which possess non-trivial thermal-momentum and thermal-winding charges. These states induce phase transitions from the Winding to the Brane regime and from the Brane to the Momentum regime, consistently with string perturbation theory. Following our previous considerations, we find some interesting classes of singularity-free string cosmological solutions:\\
 ($i$) $\{{\cal C}_{\rm String}(\tau)\} \equiv  \{{\cal W}(\tau<\tau_-)\}  \oplus 
 \{{\cal B}(\tau_-\le \tau\le \tau_+\} \oplus \{{\cal M} (\tau_+<\tau)\}$, \\
($ii$) $\{\tilde {\cal C}_{\rm String}(\tau)\} \equiv \{{\cal B}(\tau\le\tau_+)\} \oplus \{{\cal M} (\tau_+<\tau)\}$.

 The first class ($i$) generalizes the bouncing solutions of Refs\cite{FKPT,dNonSingular}, spreading the phase transition between the Winding and Momentum regimes within a finite lapse of time. During this interval, the $\sigma$-model temperature and scale factor are constant, attaining their critical values. This intermediate regime precisely corresponds to the Brane regime. In this phase, the effective potential $V(\phi)=B+Ce^{-2\phi}$ requires  $C<0$, while $B$ is fixed by the equations of motion to be equal to the critical value of the pressure, $B=P_c$. 

 The second class ($ii$) describes non-singular string cosmologies, where the Universe is in the Brane regime at very early times, $\tau \ll \tau_+$, with a flat $\sigma$-model metric, a constant maximal temperature $T_c$ and weak string interactions, $g_{\rm str}\ll 1$. The Brane effective potential $V(\phi)=B+Ce^{-2\phi}$ must have $C\ge 0$, while $B=P_c$ as in case ($i$). During the evolution, the metric remains flat up to $\tau_+$, while the string coupling grows and reaches a critical value $g^*_{\rm str}$ at $\tau_+$, where the Universe exits from the Brane regime. At later times, $\tau > \tau_+$, the Universe is in a new phase in expansion, with radiation and decreasing string coupling due to the dilaton motion. Throughout the evolution, the string coupling remains smaller than $g^*_{\rm str}$.  
 
 In the Einstein cosmological frame, the time-evolution in case ($ii$) gives rise to a new class of bouncing Universes connecting at $t=t_+$ two distinct phases. 
The initial contracting phase is characterized by a uniform time-distribution of spacelike branes and fluxes up to $t_+$. When $C>0$, the scale factor and temperature behave as in a negatively curved space with radiation. In the special $C=0$ case, the effective curvature vanishes and the Brane regime is a radiation-like era. Actually, this effective behavior arises from the non-trivial motion of the dilaton field and the definitions of the scale factor and temperature in the Einstein frame, $a_E=a_c \, g_{\rm str}^{-{2\over d-2}}$, $T_E=T_c \, g_{\rm str}^{2\over d-2}$. In the second phase, $t>t_+$, the Universe is in expansion, with thermal radiation and a running dilaton converging to a constant. 
 
In both cases ($i$) and ($ii$), the entropy is conserved during the evolution. The largeness of the entropy observed at late cosmological times implies that the size of the Universe during the Brane regime, where the temperature attains its maximal critical value, is already large. This fact guarantees the validity of the perturbative approach  during the whole cosmological evolution, and manifests the connection of the so-called entropy and oldness problems of standard Big Bang cosmology.   

 \section{Microscopic origin of the action in the Brane regime}
 \label{Micro}
 
In this Section, we investigate the microscopic origin of the effective action in the Brane regime, namely, the localized-in-time terms at the endpoints of the Brane ($\delta$-functions proportional to $\kappa_{\pm}$), as well as the Brane effective potential ($V(\phi)=B+Ce^{-2\phi}$). We show that they follow naturally from the underlying description of the stringy system at the extended symmetry point, where the temperature reaches its maximal critical value $T_c$. 

It is convenient to treat the microscopic action accordingly, when we are: \\
(1) ``Going in" and/or ``going out" of the Brane, or\\
(2) ``Being'' on the Brane. 

In case (1), the constraint $\delta (\sigma)$ is translated into localized constraints at the temporal endpoints of the brane: 
$\delta(\tau-\tau_-)$ and $\delta(\tau-\tau_+)$. In this case, the relevant microscopic action is described by a Euclidean action in $(d-1)$ dimensions. 
In Refs \cite{FKPT,dNonSingular}, this situation has been considered in great detail. The origin of a localized negative contribution to the pressure arises 
from non-trivial gradients in the spatial directions, $\langle \partial_{\hat \mu}\varphi^I\rangle\ne 0$, which the extra massless scalars at the endpoints of the brane can have,  as we review in the next Section. 
 
In case (2), the microscopic stringy system is described by an exact conformal theory at the extended symmetry point $\sigma=0$, 
during the time-interval $D\equiv (\tau_-,\tau_+)$. Here, the constraint $\delta(\sigma)$ cannot be translated in time. The effective field theory in the lapse of time  $D$ 
is well described (in the Euclidean) by a $(d+2)$-dimensional supergravity theory, where the two extra dimensions are compactified on an $SU(2)_{k=2}/ U(1)$ manifold.  

As we will show explicitly, non-trivial gauge field fluxes during the interval of time $D$ give rise to a $d$-dimensional effective potential of the form $V(\phi)=B+C\,e^{-2\phi}$, where $B, C\ge 0$. It is of the precise form to maintain in the Brane regime the constraint $\sigma=0$. We will also comment on the possibility of generating strictly negative $C$-terms and refer to Ref. \cite{cyclic} for details on this issue. 


\subsection{The localized terms at the endpoints of the Brane regime}
\label{kappa}
As we already stated, microscopically  the localized terms at the endpoints of the brane arise due to the non-trivial gradients, 
$\langle\partial_{\hat \mu}\varphi^I\rangle\ne 0$,  which the extra massless scalars present at
$\sigma=0$ acquire. These extra scalars give rise to a $(d-1)$-dimensional Euclidean action:
\be
\label{ac1}
 \S_\pm= -\int d\sigma\, d^{d-1}x \, \sqrt{g_{\bot}}\, e^{-2\phi}\,g^{\hat \mu\hat \nu}\,G_{IJ}\, \partial_{\hat \mu} \varphi^I \partial_{\hat\nu} { \varphi}^J\,\delta(\sigma),
\ee
where $\hat\mu=1,\dots,d-1$, $g_\bot=\det(g_{\hat\mu\hat\nu})$ and $G_{IJ}$ is the $\varphi$-dependent metric in the field configuration space.
The equations of motion of the scalars $\varphi^I$ take the form:
\be
\label{eomphi}
2\partial_{\hat\mu}( e^{-2\phi} \sqrt{g_\bot} ~g^{\hat\mu\hat\nu} ~G_{IJ} \partial_{\hat\nu} \varphi^J )-
e^{-2\phi} \sqrt{g_\bot} ~g^{\hat\mu\hat\nu}(\partial_IG_{KJ}) \partial_{\hat\mu} \varphi^K\partial_{\hat\nu} \varphi^J \, =0 .
\ee
They admit non-trivial solutions, which are
consistent with the homogeneity and isotropy requirements and yield the  Brane terms localized at the temporal endpoints $\tau_\pm$ \cite{dNonSingular}. 
The relevant solutions are such that the induced metric,
\be
\label{hg}
h_{\hat\mu\hat\nu}\equiv G_{IJ}~\partial_{\hat\mu}\varphi^I\,\partial_{\hat\nu}\varphi^J\,={\kappa_\pm\over d-1} ~g_{\hat\mu\hat\nu},
\ee 
is proportional to the spatial metric, where $\kappa_+$ and $\kappa_-$ are positive constants. 
When this happens, the stress tensor of the scalars is consistent with the
symmetries of the spatial metric, and therefore with homogeneity and isotropy. 
Moreover, the actions $\S_+$ and $\S_-$  take the familiar form of the Nambu-Goto action for branes,
\begin{eqnarray}
 \nonumber \S_\pm\!\!\!&=&\!\!\!-\kappa_\pm\int d^d x\,e^{-2\phi}\, \sqrt{g_\bot}~ \delta(\tau-\tau_\pm)\\ 
 \!\!\!&=&\!\!\!- (d-1)^{d-1\over 2}\kappa_\pm^{3-d\over 2}\int d^dx\, e^{-2\phi}\, \sqrt{\det\left(G_{IJ} \partial_{\hat\mu} \varphi^I\partial_{\hat\nu} \varphi^J\right)}~ \delta(\tau-\tau_\pm)\,.
\end{eqnarray}
 Thus, $\kappa_\pm$ are interpreted as brane tensions. In the case $k=0$, the isotropic embeddings  $x^{\hat\mu} \rightarrow \varphi^I(x^{\hat\nu})$ are given by
\be
\label{FlatEmb}
\partial_{\hat \mu} \varphi^I=\sqrt{{2\kappa_\pm \over d-1}}\,a_c\,\delta^I_{\hat\mu}\,,
\ee 
so that both $g_{\hat \mu\hat\nu}$ and $G_{\hat \mu\hat\nu}$ are flat at $\tau=\tau_\pm$. 
As has been explained in Ref.\cite{dNonSingular}, the realization of this isotropic embedding imposes a constraint on the dimensionality 
of space-time, namely  $d\le 6$. 

\subsection{The Brane effective potential $V$ from field strength fluxes}

At the extended symmetry point (where the Brane is located), 
the $U(1)$ Euclidean time circle is extended to an $SU(2)_{k=2}$ manifold, 
so that the target space is naturally $(d+2)$-dimensional. Thus, instead of a ``naive'' field theory in $d$ space-time dimensions, the Brane effective action during the time-interval $D\equiv (\tau_-, \,\tau_+)$ can be written as
\be
\label{extra2}
\S_D=-\int  d^{d+2} x \, \sqrt{g} \, \Big\{  {\cal A} (\tilde \phi, \Phi^\alpha)\, (\partial_M\Phi^\beta)^2\, +\, \B(\tilde\phi, \Phi^\alpha) \, (F_{MN})_L^2\,+\, \C(\tilde\phi, \Phi^\alpha) \, (F_{MN})_R^2 +\cdots \Big\} ,
\ee
where $\tilde\phi$ is the dilaton in $(d+2)$-dimensions, the $\Phi^\alpha$'s are the internal massless scalars, $(F_{MN})_L$ is the field strength of the graviphotons, $(F_{MN})_R$ is the field strength of the matter gauge bosons. The ellipses denote the contributions of the metric, antisymmetric tensor and fermionic fields of the effective {\it gauged supergravity theory} \cite{deRoo} in $(d+2)$ dimensions. In the presence of non-trivial fluxes arising either from the gradients of the scalars or from the gauge field strenghts, an effective potential of the following form is generated for the dilaton in $(d+2)$ dimensions,
\be
V(\phi)= {\cal \tilde  A}(\tilde \phi,\Phi^\alpha)+\tilde  \B(\tilde \phi, \Phi^\alpha) +\tilde  \C({\tilde \phi, \Phi^\alpha}).
\ee 
This form will persist when the theory is reduced to $d$ dimensions, since the two extra compact dimensions of the $SU(2)_{k=2}/U(1)$ manifold have a
finite volume of order 1 in string units. 
In order to derive the dilaton dependence of each individual term, we will utilize two  facts: 

$\bullet$ In the $\sigma$-model frame, the dilaton dependencies of the terms ${\tilde  {\cal A}}$, $\tilde  \B$ and $\tilde  \C$ is independent of the dimensionality of space-time. 

$\bullet$  The dilaton dependence of the gauge kinetic terms is dictated by the structure of
the $\N_4=4$ supergravity theory, (gauged or not), in $d+2=4$  space-time dimensions \cite{deRoo}.
 
\noindent Combining these two properties, we will derive unambiguously how the three individual flux terms $\tilde {\cal A}$,  $\tilde  \B$ and $\tilde  \C$ depend on the $d$-dimensional dilaton field $\phi$,
\be
\label{ft}
\tilde {\cal A}=A(\Phi^\alpha) e^{-2\phi}~,~~~\tilde \B=B(\Phi^\alpha)~, ~~~\tilde \C =\tilde C(\Phi^\alpha)e^{-2\phi} .
\ee

Let us discuss in more detail the origin of the action (\ref{extra2}). The starting point is the type II superstring, with $n_c=10-d$ internal directions compactified on a  torus $T^{n_c}$.  The coupling (\`a la Scherck-Shwarz \cite{SS}) of at least one of the compact dimensions to right-moving R-symmetry charges breaks the initial $\N_4$=(4,4) 
supersymmetry to a left-moving $\N_4$=(4,0) supersymmetry. Thus, at least one of the $n_c$ directions, say $x^9$ as in the model of Ref. \cite{FKPT}, is compactified at the fermionic point and implies an $U(1)_{9R}\to [SU(2)_{9R}]_{k=2}$ extended symmetry. The temperature coupling to the left-moving sector is introduced in the compact Euclidean direction  $x^0$. At the fermionic point associated to the Brane regime,  the $U(1)_{0L}$ symmetry of the temporal circle is extended to $[SU(2)_{0L}]_{k=2}$. Reorganizing the symmetries as,
\begin{align}
\nonumber\bigg(SU(2)_{0L}\otimes U(1)_{0R}\bigg)\times& \bigg(U(1)_{9L}\otimes SU(2)_{9R}\bigg)=\\
&\bigg(SU(2)_{0L}\otimes \Big[ U(1)_{0R}\times {SU(2)_{9R}\over U(1)_{9R}}\Big]\bigg)\times \bigg(U(1)_{9L}\otimes U(1)_{9R}\bigg),
\end{align}
the  temporal cycle $S^1$ is extended to a  3-dimensional sphere $S^3$. 
This justifies why we considered the $\N_4=(4,0)$ supergravity theory in $(d+2)$ dimensions in Eq. (\ref{extra2}). The introduction of the temperature coupling in the left-moving sector
can be interpreted (in the Euclidean) as a gauging of the $\N_4=(4,0)$ supergravity in one less non-compact spatial dimension, but with 2 extra compact directions arising when the temperature is critical. 

On the contrary, ``outside the Brane" ($T<T_c$), the geometrical interpretation in $(d+2)$ dimensions is not valid anymore. This statement is very crucial in what follows, since the spatial derivatives appearing in the gradients of the scalars and gauge kinetic terms in Eq. (\ref{extra2}) will be taken in these two extra  directions.  As a result, the scalar and gauge fluxes considered above do not exist outside the Brane regime.

 We are now in a position to determine the induced functional forms of the individual flux terms in Eq. (\ref{ft}). The easiest one is that of the scalar fields. The definition of the dilaton $\tilde \phi$ in $(d+2)$ dimensions absorbs the volume factor $\sqrt{{\rm det}\,g_{ij}}$ of the $n_c$  internal compact dimensions.  Moreover, there is no dependence in  the two extra directions ($z,\bar z$), whose volume is finite and of order 1. Combining these two facts, we find in terms of the $d$-dimensional dilaton field that ${\cal \tilde A}=A\, e^{-2\phi}$, where A is a constant which absorbs the volume factor $\sqrt{{\rm det}\,g_{z \bar z}}$.
 
We now come to the $\B$ and $\C$ terms. Both are generated by some ``gauge" fluxes in the spatial directions $(M,N)=(z,\bar z)$.  
Restricting for the moment to the $d+2=4$ dimensional case and according to the ``gauged'' $\N_4=4$ supergravity structure \cite{deRoo}, 
we conclude that there are two possible dilaton dressings multiplying the gauge kinetic terms (in the Einstein-frame). 
This is due to the fact that the gauge bosons in the supergravity multiplet (left-moving graviphotons) and the gauge bosons in the matter supermultiplets 
transform non-trivially under the $SU(1,1)$ duality symmetry of the dilaton-axion field. 
The latter parametrizes an  $[SU(1,1)/U(1)]_S$ manifold, where $S=e^{-2\tilde \phi} +i\chi$. (The axion field $\chi$ is the dual of the two-index 
antisymmetric tensor $B_{MN}$). The $SU(1,1)$ symmetry implies a different coupling of $S$ to the graviphotons and matter gauge fields:
\be
\label{N=4Restriction}
\B\,(F^2_{MN})_{\rm graviphotons} +\C\, e^{-2\tilde \phi}\,(F^2_{MN})_{\rm matter\mbox{\footnotesize -}gauge\mbox{\footnotesize -}fields}.
\ee
In four dimensions, this expression takes an identical form, when written in the $\sigma$-model frame (due to fact that $\sqrt{{\rm det}\, g_{MN}}\; g^{PP'}g^{QQ'}$ is invariant under a conformal rescaling of the metric). The above constraints of $\N_4=4$ gauge supergravity in $d+2=4$ dimensions turn out to be very 
efficient. Indeed, as announced before, the dilaton dependencies of the $\B$ and $\C$ terms in the $\sigma$-model frame remain the same in any dimension. Furthermore, assuming non-trivial fluxes in the $(M,N)=(z,\bar z)$ extra dimensions, 
and absorbing the volume factor of the $n_c$ compact directions  $\sqrt{{\rm det}\,g_{ij}}$ in the definition of the $d$-dimensional dilaton $\phi$, we find that
\be
\tilde \C(\tilde \phi,\Phi^\alpha) =\tilde C\, e^{-2\phi}~,~~~\tilde\B(\tilde \phi, \Phi^\alpha) =B({\cal U})\, ,
\ee
where $\tilde C$ is a constant, independent of the internal moduli fields $\Phi^\alpha$. As was the case for the $\tilde {\cal A}$-term, there is no dependence in the $(z, \bar z)$ extra dimensions, whose volume factor $\sqrt{{\rm det}\,g_{z \bar z}}$ is of order 1 and absorbed in the definitions of $\tilde C$ and $B$.
Since the $\tilde \B$-term has no dilaton dependence, the $\Phi^\alpha$-dependent internal volume cannot be absorbed in the definition of the $d$-dimensional dilaton. 
Thus, although this term is independent of the dilaton, it does depend on the volume $\cal U$ of the internal torus $T^{n_c}$. 

We conclude, that the Brane effective potential for the dilaton $\phi$ in $d$ dimensions, which is induced by the fluxes, takes the form
\be
V(\phi) = B({\cal U}) + C\,e^{-2\phi},
\ee 
where $C=A+ \tilde C$ is a constant and $\cal U$ is the volume of the internal $(10-d)$-dimensional  space. 
According to Section \ref{BraneDistributionV}, the dynamical requirement of being on the brane fixes the volume ${\cal U}$ such that $B({\cal U})=P_c\,$. 

Notice that both the $B$ and $C$ terms are positive, once their  microscopic origin is due to the above considered fluxes. 
This seams to exclude the $C<0$ case studied in Section \ref{EffectiveActions}. A natural question arising at this point, 
is the possibility of generating an effective $C<0$ from different microscopic considerations. 
The answer to this question is affirmative. Indeed, such a negative term can arise, when one considers an underlying conformal field theory 
with a central charge deficit at genus-$0$. This can be easily realized in four spacetime dimensions, when the 3-dimensional space is $S^3$ rather than flat. 
The effective $C$ in this case
is proportional to the central charge deficit, $C\propto \delta \hat c =- {4\over (k+2)}$, where $k$ is the
level of $SU(2)_k\sim S^3$. 
We plan to analyze this possibility elsewhere \cite{cyclic}.

\section{Conclusions}
\label{conclu}

In this work, we investigated cosmological consequences of the recently discovered stringy gluing mechanism between different string effective field theories. This mechanism is induced by the appearance of extra massless string states possessing non-trivial thermal-momentum and thermal-winding charges, and localized at cosmological times when the temperature reaches its maximal critical value.  

The effective field theory regime with constant $\sigma$-model frame  temperature equal to its maximal value ($T=T_c$) admits a natural ``brane interpretation", with tensions $\kappa_{\pm}$ given by the non-trivial gradients $\langle\partial_{\hat\mu}\varphi^I\rangle\ne 0$, which the extra massless scalars acquire at the endpoints of the Brane regime.  
Non-trivial gauge fluxes on the Brane give rise to an effective dilaton potential, $V(\phi) =B({\cal U}) + C\,e^{-2\phi}$,
which turns out to be compatible with the requirement of ``being on" the Brane, with constant temperature equal to its maximal value during the time interval 
$\tau_-<\tau<\tau_+$. The $B$-term is proportional to the volume of the internal space, and fixed to
be equal to the critical pressure at $T_c$, $B({\cal U})=P_c$, by the dilaton and gravitational field equations of motion. The coefficient $C$ turns out to be constant and behaves like an effective central charge ``deficit/benefit" in the underlying two-dimensional worldsheet superconformal theory. 

When $C\ge 0$, the earliest endpoint of the brane 
is forced to be at $\tau_-=-\infty$.  This fact gives rise to a new class of non-singular string cosmologies:
$\{{\cal C}_{String}(\tau)\} \equiv \{{\cal B}(\tau\le\tau_+)\} \oplus \{{\cal M} (\tau_+<\tau)\}$, 
where the Universe at very early times stays in the Brane regime, corresponding to an exact worldsheet conformal field theory. It is characterized by a constant maximal $\sigma$-model temperature $T_c$, flat $\sigma$-model metric and by super weak string interactions, $g_{\rm str}\ll 1$. During the evolution, the temperature and the metric remain stationary up to $\tau_+$, while the string coupling grows and reaches a critical value $g^*_{\rm str}$  at $\tau_+$.  At later times, $\tau> \tau_+$, the Universe enters in the Momentum phase, $\{{\cal M}(\tau>\tau_+)\}$, which is described  by a radiation dominated expansion and decreasing string coupling converging to a constant.

In the Einstein cosmological frame, the time-evolutions of the  $\{{\cal C}_{\rm String}(\tau)\}$ $\equiv  
 \{{\cal B}(\tau\le \tau_+)\} \oplus \{{\cal M} (\tau_+<\tau)\}$
cosmologies, give rise to a new class of  bouncing Universes connecting at $t=t_+$ two distinct phases. The initial  one is contracting and characterized by a scale factor and temperature evolving as in a  negatively curved space filled with radiation. At later times, the Universe enters in an expanding thermal phase with a decreasing dilaton, which is asymptotic to a constant. 

When $C<0$, the Brane regime has two endpoints, $\tau_\pm$. In this case, the Universe has three characteristic phases:
 $\{{\cal C}_{\rm String}(\tau)\} \equiv  \{{\cal W}(\tau<\tau_-)\}  \oplus 
 \{{\cal B}(\tau_-\le \tau\le \tau_+)\} \oplus \{{\cal M} (\tau_+<\tau)\}$.
Namely, for $\tau<\tau_-$, the Universe is in a contracting Winding regime, up to the time $\tau_-$ when the temperature and the string coupling reach their maximal critical values, $T_c$ and $g^*_{\rm str}$.  
At $\tau_-$, the Universe enters in the Brane regime, which lasts for a finite amount of time, having constant temperature $T=T_c$. The string coupling decreases  
reaching a minimal value $g_{\rm str}^{\rm min}$ at $\tau_{\rm min}/2$, and then grows again reaching its maximal value at $\tau_+$, the second endpoint of the Brane.  At $\tau_+$, the Universe enters in the Momentum expanding regime, with decreasing temperature and string coupling, the latter being asymptotic to a constant.  
The $C<0$ solutions with an intermediate  Brane regime ($\tau_-<\tau<\tau_+$) extend the non-singular bouncing cosmologies of Refs \cite{FKPT,dNonSingular}, where the Brane regime collapses to a single instant in time. 

The cosmological solutions described in this work remain perturbative throughout the evolution, provided that the critical value of the string coupling at the endpoints of the brane is sufficiently small. An important property is that the entropy is conserved during  the whole evolution, with the critical value of the scale factor at the endpoints of the Brane regime being determined by  it and the maximal critical temperature. This class of bouncing cosmologies as well as the those of Refs \cite{FKPT,dNonSingular} are the first higher dimensional examples, where both the Hagedorn instability as well as the classical Big Bang singularity are successfully resolved, remaining in a perturbative regime throughout the evolution. 
 
 We have presented spatially flat solutions ($k=0$). However, there are other interesting possibilities with $k\ne 0$. In particular, when the stringy gluing mechanism is applied to the $k>0$ case, it is possible to realize a cyclic closed cosmology, where the apparent Big Bang and Big Crunch singularities are resolved by the appearance of a periodic array of branes. This work is currently under progress \cite{cyclic}. It would be interesting to apply the stringy gluing mechanism in other physically interesting problems, where paradoxes related to spacelike singularities appear, such as in the interior of black hole geometries.    

Having at our disposal exact cosmological solutions, we can explicitly calculate the spectrum of fluctuations at early times, say at the time locations of the branes, determine their propagation at later cosmological times and compare them to the current and future observational data. This is possible since we have analytical control on the theory describing the brane. This work is currently under progress \cite{BKPPT}.

  
 \section*{Acknowledgement}
 
 We are grateful to C. Bachas, R. Brandenberger, D. L\"ust, S. Patil, J. Troost and especially I. Florakis for fruitful discussions. C.K. and H.P. would like to thank the University of Cyprus for hospitality. N.T. and H.P. acknowledge the Laboratoire de Physique Th\'eorique of Ecole Normale Sup\'erieure for hospitality. N.T. would like to thank the Centre de Physique Th\'eorique of Ecole Polytechnique for hospitality. The work of C.K. and H.P. is partially supported by the ANR 05-BLAN-NT09-573739 and IRSES-UNIFY contracts. The work of C.K., H.P. and N.T. is also supported by the CEFIPRA/IFCPAR 4104-2 project and a PICS France/Cyprus. The work of H.P. is partially supported by the EU contracts PITN GA-2009-237920, ERC-AG-226371 and PICS  France/Greece, France/USA.

\vspace{.4cm}

\end{document}